# Force Induced Unzipping of DNA with Long Range Correlated Noise

Pui-Man Lam and Yi Zhen
Physics Department, Southern University
Baton Rouge, Louisiana 70813

**Abstract.** We derive and solve a Fokker-Planck equation for the stationary distribution of the free energy, in a model of unzipping of double-stranded DNA under external force. The autocorrelation function of the random DNA sequence can be a general form, including long range correlations. In the case of Orstein-Uhlenbeck noise, characterized by a finite correlation length, our result reduces to the exact result of Allahverdyan et al, with the average number of unzipped base pairs going as $<X> \sim 1/f^2$ in the white noise limit, where $f$ is the deviation from the critical force. In the case of long range correlated noise, where the integrated autocorrelation is divergent, we find that $<X>$ is finite at $f=0$, with its value decreasing as the correlations become longer range. This shows that long range correlations actually stabilize the DNA sequence against unzipping. Our result is also in agreement with the findings of Allahverdyan et al, obtained using numerical generation of the long range correlated noise.

PAC number(s): 82.37.Rs, 87.14.Gg, 87.15.Aa

## I. Introduction

In the past two decades, micromanipulation techniques have become important tools in the repertoire of biophysicists and structural biologists, complementing more traditional scattering and spectroscopic measurements. In the DNA molecule, the two individual strands are bonded by hydrogen bonds while they themselves are constructed by much stronger covalent bonds. In addition, the two double strands are wrapped around each other in a double helix structure. For simplicity, this last aspect of the DNA structure will be neglected in our study. Each single strand is a polymer formed from nucleotides which can be of two types: purines, consisting of adenine (A) and guanine (G) and pyrimidines, consisting of cytosine (C) and thymine (T). The hydrogen bonds between the two opposite strands can only be of two types, either A-T bases or G-C bases, with different formation energies. The GC base pairs are made of three hydrogen bonds while AT base pairs are made of two hydrogen bonds only. These base pairs, since they are hydrophobic, are located at the core of the double helix. The polymerase, whose function is to read the genetic code encoded in the DNA, must first unzip the two strands in the DNA in order to get to it. This makes force induced unzipping of DNA by an external force an important mechanism in the functioning of all living organisms. Force induced unzipping of DNA has been actively investigated in the last decade [1-9]. For a very recent thorough review on biomolecules under mechanical force, see the article by Kumar and Li [10].

It is known that the concentration of AT and GC base pairs is approximately equal, especially in higher organisms [11]. The difference between the AT and GC formation energies of one hydrogen bond is of the same order as the average formation energy of 2.5 hydrogen bonds. For certain bulk properties, this difference may not be relevant and DNA can be considered as a homogeneous base sequence. However for situations where the unzipping energy is of the order of the formation energy, the heterogeneity of the base sequence becomes relevant. Lubensky and Nelson [4] took the first step in this direction. They showed that for a homogeneous sequence, the number of unzipped base pairs <X> diverges as $<X> \sim (\Im_c - \Im)^{-1}$ for external force $\Im$ near the critical for force $\Im_c$, while for a heterogeneous sequence with short range white noise correlation, it diverges as $<X> \sim (\Im_c - \Im)^{-2}$.

It is known that DNA sequences in fact display long range correlations[12-15], both in the non-coding (intron) and coding regions of the DNA : two base pairs separated by thousands of pairs appear to be statistically correlated. Despite their ubiquity, the biological reason for these long range correlations are largely unexplored. For long range correlation with the correlation function $<\boldsymbol{h}(m)\boldsymbol{h}(m')> \sim |m-m'|^{\boldsymbol{a}}$, where $\boldsymbol{h}(m)$ is the noise or deviation of the binding energy from its average value at the base-pair position m along the sequence, $\boldsymbol{a} < 1$, Lubensky and Nelson [4] had predicted, using heuristic arguments, that the divergence of <X> near the critical unzipping force should go as $<X> \sim (\Im_c - \Im)^{-2/\boldsymbol{a}}$. This is a stronger divergence than the short range, white noise case.

Allahverdyan et al [7] studied the unzipping of DNA with correlated base sequence using a somewhat simplified model of Lubensky and Nelson. For the case of finite range correlation characterized by a finite correlation length, they derived and solved a Fokker-Planck equation for the distribution, from which they calculated the average number of unzipped base pairs <X>. The white noise limit could be obtained by taking the zero correlation length limit. In this limit, they recovered the result $<X> \sim (\Im_c - \Im)^{-2}$ of Lubensky and Nelson [4]. For the case of long range correlated noise, they could not derive a Fokker-Planck equation. Rather, the long range correlated noise with the correct behavior of the correlation function was numerically generated and used to calculate the average number of unzipped base pairs <X>. Contrary to the prediction of Lubensky and Nelson, they found that long range correlations actually stabilize the DNA against unzipping. Here we study the force induced unzipping of DNA with both finite range and long range correlations using an approximate Fokker-Planck. For finite range correlations we reproduce exactly the results of Allahverdyan et al. For long range correlations, we also find that the DNA is more stable compared to the short range correlated case. In section II we present the details of the model. In section III we present the derivation of the Fokker-Planck for noise autocorrelation function of a general form from which we can obtain the average free energy and the average number of unzipped base pairs. In section IV we present the solution of the stationary Fokker-Planck equation for the case of Orstein-Uhlenbeck noise, where the autocorrelation function of the noise $K(t)$ has a finite correlation length. In this case we show that the stationary probability distribution function for the free energy reduces to the exact result of Allahverdyan et al [7]. In section V, we present the solution for the case of a long range correlated noise. We find that long range correlations actually stabilize the DNA against unzipping. Our findings corroborate the findings of Allahdyan et al. [7] obtained using numerical simulation and stand in contradiction to the predictions of Lubensky and Nelson [4]. There has been very little work on the effect of long rang correlation on DNA unzipping besides the references [4] and [7]. Our Fokker-Planck equation approach may provide a new perspective. Section VI is a conclusion.

**II. The Model**

We will now review the model studied by Allahverdyan et al. The physical basis of the model has been described in detailed in their paper and will not be repeated here. In the next section we will concentrate on the derivation of a Fokker-Planck equation that can be applied to both short range and long range correlated noise in the DNA sequence. The DNA lies along the x-axis between $x = a$ and $x = L$. The base pairs are located at points $x_i$, $a \leq x_i \leq L$, $i = 1,2...M$. They can be in one of two states: bound or disconnected. Disconnected base pair at point $x_i$ contributes a binding energy $f(x_i)$, while bound pairs contribute nothing. The binding energy $f(x_i)$ is a random quantity with an average $<f>$:

$$f(x_i) = <f> + h(x_i) \qquad (1)$$

where $h(x_i)$ is a random deviation from the average value at point $x_i$. An external force $\Im$ is acting on the left end $x = a$, pulling apart the two strands. If a bond at point $x_i$ is broken, then all base pairs with $j < i$ are broken as well.

The Hamiltonian is given by

$$H(x) = -\Im x + \sum_{i=1}^{x} f(x_i) = -\Im x + \sum_{i=1}^{x} [<f> + h(x_i)]$$

$$= (<f> - \Im)x + \sum_{i=1}^{x} h(x_i) \equiv fx + \sum_{i=1}^{x} h(x_i) \quad , \qquad (2)$$

where $x$ is the number of broken base pairs and

$$f \equiv <f> - \Im \qquad (3)$$

Here $\Im$ denotes both the force and the force multiplied by the base-pair separation, which is taken as unity. The units in Eqn. (2) and (3) may look strange. This is because the base-pair separation is taken as unity. In this units, force and energy has the same notation. Similarly, $x$, $x_i$ can either be numbers or distance in units of base-pair separation. In this Hamiltonian, the two unzipped single strands exert no restoring force.

This is a simplification of the model studied by Lubensky and Nelson. Of course in both models, the long range, excluded volume interaction of the polymers has been neglected. However, using this model, Allahverdyan et al reproduced the results of Lubensky and Nelson on the divergence of the average number of unzipped base pairs, when the external force is close to the critical value, for the case when $h(x_i)$ are short range correlated. This shows that this model can also be used to study the effect of long range correlation on the average number of unzipped base pairs.

In the continuum limit, the Hamilton becomes,

$$H(x) = (x-a)f + \int_a^x ds\, h(s) \qquad (4)$$

From this we can calculate the partition function and the free energy.

$$Z = \int_a^L dx\, e^{-bH(x)}, \quad F = -T \ln Z, \qquad (5)$$

with $b = 1/T$ the inverse temperature and the Boltzmann's constant $k_B \equiv 1$.

The order parameter is the number of broken base pairs $X$:

$$X = \partial_f F, \quad <X> = \partial_f <F>. \qquad (6)$$

It remains to specify the properties of the noise $h$. We assume an autocorrelation function of the noise of the form

$$K(t-t') \equiv <h(t)h(t')>, \quad K(t) = K(-t). \qquad (7)$$

Depending on the behavior of $K(t)$ for large $t$, two cases are distinguished: finite range correlated situation and long range correlated situation.

In the finite range situation, the total intensity of the noise is finite:

$$D = \int_0^\infty ds\, K(s). \qquad (8)$$

In particular, the white noise case

$$K(t) = D d(t) \qquad (9)$$

describes completely uncorrelated noise.

The Ornstein-Uhlenbeck noise is characterized by a finite correlation length $t$:

$$K(t) = \frac{D}{t} e^{-|t|/t}. \qquad (10)$$

Power law correlated noise is given by

$$K(t) \sim |t|^{-d}, \quad |t| \geq 1, \quad d > 1. \tag{11}$$

Long range correlated noise is characterized by

$$K(t) \sim s|t|^{-a}, |t| \geq 1, 0 < a < 1. \tag{12}$$

Here we take $K(t)$ to be

$$K(t) = \begin{cases} s, |t| \leq 1 \\ s|t|^{-a}, |t| > 1, a < 1 \end{cases}. \tag{13}$$

A Langevin equation can be obtained by differentiating Eqn. (5) with respect to $a$

$$\frac{dZ}{da} = -e^{-bH(a)} - \int_a^L dx \, b e^{-bH(x)} \frac{\partial H}{\partial a}$$

$$= -\exp(0) + b \int_a^L e^{-bH(x)} [f + h(a)]. \tag{14}$$

With the substitution $t = -a$, this can be rewritten as

$$\frac{dZ}{dt} = 1 - bfZ - bh(t)Z, \quad -L < t < 0 \tag{15}$$

In terms of the free energy $F$, this can be written as

$$\frac{dF}{dt} + Te^{bF} - f = h(t), \quad -L < t < 0. \tag{16}$$

### III. Derivation of a Fokker-Planck Equation

In this section we will derive a Fokker-Planck equation for the probability distribution $P(F,t)$ of the free energy at time $t$, corresponding to the Langevin equation (16). We will follow the functional integral method of Fox [16]. Any Gaussian noise has the functional form

$$P[h] = \Re \exp[-\frac{1}{2} \int ds \int ds' h(s) h(s') C(s - s')], \tag{17}$$

where $\Re$ is a normalization factor

$$\mathfrak{R}^{-1} = \int D\mathbf{h} \exp[-\frac{1}{2}\int ds \int ds' \mathbf{h}(s)\mathbf{h}(s')C(s-s')] . \qquad (18)$$

The function $C(s)$ is the functional inverse of the autocorrelation function $K(t)$ defined in Eqn. (7) i.e.

$$\int ds C(t-s)K(s-s') = \mathbf{d}(t-s) . \qquad (19)$$

This can be shown by the following analysis (as suggested by an anonymous referee).

$$0 = \mathfrak{R}\int D\mathbf{h} \frac{d}{d\mathbf{h}(t)}\left\{\mathbf{h}(s)\exp[-\frac{1}{2}\int ds'\int ds''\mathbf{h}(s')\mathbf{h}(s'')C(s'-s'')]\right\}$$

$$= \mathfrak{R}\int D\mathbf{h}\left\{\mathbf{d}(t-s) - \mathbf{h}(s)\int ds' ds''\mathbf{d}(t-s')C(s'-s'')\mathbf{h}(s'')\right\}\exp[-\frac{1}{2}\int ds'\int ds''\mathbf{h}(s')\mathbf{h}(s'')C(s'-s'')]$$

$$= \mathbf{d}(t-s) - \int ds'' C(t-s'')K(s-s'')$$

The probability for having free energy $F$ at time $t$ is given by

$$P(F,t) = \int D\mathbf{h} P[\mathbf{h}(t)]\mathbf{d}(F - F[\mathbf{h},t]) . \qquad (20)$$

The time derivative of this is given by

$$\frac{dP}{dt} = \int D\mathbf{h} P[\mathbf{h}(t)][-\frac{\partial}{\partial F}\mathbf{d}(F - F[\mathbf{h},t])]\frac{dF[\mathbf{h},t]}{dt} \qquad (21)$$

Using Eqn. (16), this can be written as

$$\frac{dP}{dt} = -\frac{\partial}{\partial F}(f - Te^{F/T})\int D\mathbf{h} P[\mathbf{h}]\mathbf{d}(F - F[\mathbf{h},t]) - \frac{\partial}{\partial F}\int D\mathbf{h} P[\mathbf{h}]\mathbf{d}(F - F[\mathbf{h},t])\mathbf{h}(t) . \qquad (22)$$

From the form of $P[\mathbf{h}]$ in (17) and equation (19) one can easily show that

$$\mathbf{h}(t)P[\mathbf{h}] = -\int \frac{dP[\mathbf{h}]}{d\mathbf{h}(u)}K(u-t)du \qquad (23)$$

Substituting this into the second term in Eqn. (22) one has

$$-\frac{\partial}{\partial F}\int D\mathbf{h}\mathbf{d}(F - F[\mathbf{h},t])P[\mathbf{h}]\mathbf{h}(t)$$

$$= \frac{\partial}{\partial F} \int K(u-t)du \int D\mathbf{h} \frac{dP[\mathbf{h}]}{d\mathbf{h}(u)} \mathbf{d}(F - F[\mathbf{h},t])$$

$$= -\frac{\partial}{\partial F} \int K(u-t)du \int D\mathbf{h} P[\mathbf{h}] \frac{d}{d\mathbf{h}(u)} \mathbf{d}(F - F[\mathbf{h},t])$$

$$= \frac{\partial}{\partial F} \int K(u-t)du \int D\mathbf{h} P[\mathbf{h}] \frac{d}{dF} \mathbf{d}(F - F[\mathbf{h},t]) \frac{dF[\mathbf{h},t]}{d\mathbf{h}(u)}$$

$$= \frac{\partial^2}{\partial F^2} \int_{-L}^{0} ds K(t-s) < \mathbf{d}(F - F[\mathbf{h},t]) \frac{dF[\mathbf{h},t]}{d\mathbf{h}(s)} > \qquad (24)$$

where the second equality was obtained by a functional integration by parts.

The functional derivative $\frac{dF[\mathbf{h},t]}{d\mathbf{h}(s)}$ involved in the last equality in Eqn. (24) can be obtained using the Langevin equation (16).

$$\frac{dF[\mathbf{h},t]}{d\mathbf{h}(s)} = \mathbf{q}(t-s) \exp[\int_t^s e^{F[\mathbf{h},t']/T} dt'] \qquad (25)$$

where $\mathbf{q}(t)$ is the Heavyside step-function. Using this in the last equality in Eqn. (24) one can write the second term in Eqn. (22) as

$$-\frac{\partial}{\partial F} \int D\mathbf{h} \mathbf{d}(F - F[\mathbf{h},t]) P[\mathbf{h}] \mathbf{h}(t)$$

$$= \frac{\partial^2}{\partial F^2} \int_{-L}^{0} ds K(t-s) < \mathbf{d}(F - F[\mathbf{h},t]) \mathbf{q}(t-s) \exp[\int_t^s e^{F[\mathbf{h},t']/T} dt'] > \qquad (26)$$

We now make the following approximation to the integral on the right hand side of Eqn. (26):

$$\int_{-L}^{0} ds K(t-s) < \mathbf{d}(F - F[\mathbf{h},t]) \mathbf{q}(t-s) \exp[\int_t^s e^{F[\mathbf{h},t']/T} dt'] >$$

$$\approx \int_{-L}^{t} ds K(t-s) P(F,t) \exp[-e^{F/T}(t-s)]$$

$$= P(F,t) \int_{-L}^{t} ds K(t-s) \exp[-e^{F/T}(t-s)] \qquad (27)$$

This approximation consists in replacing $F[\mathbf{h},t]$ within the expectation value by the constant $F$. The justification for this is the existence of the Dirac delta function $\mathbf{d}(F - F[\mathbf{h},t])$. Using the property $K(t) = K(-t)$, this can be written as $P(F,t) \int_0^{L+t} ds K(s) \exp[-e^{F/T} s]$. In the limit $L \to \infty$, this becomes $P(F,t) \int_0^{\infty} ds K(s) \exp[-e^{F/T} s] \equiv P(F,t) G(F)$, with

$$G(F) = \int_0^\infty ds K(s) \exp[-e^{F/T} s]. \qquad (28)$$

With this approximation, the second term in Eqn. (22) can finally be written as

$$-\frac{\partial}{\partial F} \int D\mathbf{h} d(F - F[\mathbf{h},t]) P[\mathbf{h}] h(t) \approx \frac{\partial^2}{\partial F^2} G(F) P(F,t) \qquad (29)$$

with $G(F)$ given in Eqn. (28). Eqn. (22) can now be written as

$$\frac{\partial P}{\partial t} = -\frac{\partial}{\partial F}[(f - Te^{F/T}) P(F,t)] + \frac{\partial^2}{\partial F^2}[G(F) P(F,t)] \qquad (30)$$

which is the Fokker-Planck equation for $P(F,t)$.

The stationary distribution is given by $\frac{\partial P}{\partial t} = 0$, or

$$\frac{\partial}{\partial F} \{\frac{\partial}{\partial F}[G(F) P(F)] - (f - Te^{F/T}) P(F)\} = 0 \qquad (31)$$

This equation can be solved as

$$P(F) = \frac{const}{G(F)} \exp[\int_0^F dF' \frac{f - Te^{F'/T}}{G(F')}] \qquad (32)$$

Knowing $P(F)$, we can then calculate the average free energy $<F>$ as

$$<F> = \int_{-\infty}^\infty dF P(F) F = \frac{\int_{-\infty}^\infty dF \frac{F}{G(F)} \exp[\int_0^F \frac{f - Te^{F'/T}}{G(F')} dF']}{\int_{-\infty}^\infty dF \exp[\int_0^F \frac{f - Te^{F'/T}}{G(F')} dF']/G(F)} \qquad (33)$$

and the average number of unzipped base pairs $<X>$ as

$$<X> = \partial_f <F> \qquad (34)$$

In the following sections we will present solutions of the stationary Fokker-Planck equation for the cases of Orstein-Uhlenbeck noise, frozen noise and long range correlated noise.

**IV. Orstein-Uhlenbeck Noise**

The autocorrelation function for Orstein-Uhlenbeck noise is given in Eqn. (10). Substituting this function into Eqn. (28) for the function $G(F)$ we find

$$G(F) = \frac{D}{1 + te^{F/T}} \qquad (35)$$

Using this in Eqn. (32) for the stationary distribution of the free energy, we find

$$P(F) \sim [1 + te^{F/T}] \exp\left\{\frac{1}{D}\left[fF - T(T - ft)e^{F/T} - \frac{T^2 t}{2} e^{2F/T}\right]\right\} \qquad (36)$$

This agrees exactly with the result Eqn. (34) of Allahverdyan at al [7], which is an exact result. They have shown in their paper that in the white noise limit, where the correlation length $t \to 0$, this distribution of the free energy would lead to the average number of unzipped base pairs as

$$<X> \sim \frac{1}{f^2} \qquad (37)$$

**V. Long Range Correlated Noise**

We will now consider the case when the noise autocorrelation function $K(t)$ is long range and is given by Eqn. (13), with $a < 1$. Substituting Eqn. (13) into Eqn. (28), we find

$$G(F) = -se^{-F/T}[1 - \exp(-e^{F/T})] + s(e^{-F/T})^{1-a}\Gamma(1-a, e^{F/T}) \qquad (38)$$

Where $\Gamma(a, x)$ is the incomplete $\Gamma$-function

$$\Gamma(a, x) = \int_x^\infty dt\, t^{a-1} e^{-t} \qquad (39)$$

In order to use Eqns. (32) and (38) to calculate the stationary free energy distribution, we need to calculate the integral

$$\int_{F_0}^F \frac{f - Te^{F'/T}}{G(F')} dF' = \frac{1}{s}\int_{F_0}^F e^{F'/T} \frac{(f - Te^{F'/T})}{1 - \exp(-e^{F'/T}) + e^{aF'/T}\Gamma(1-a, e^{F'/T})} dF' \qquad (40)$$

With the substitution $x' = e^{F'/T}$, this integral becomes

$$\int_{F_0}^{F} \frac{f - Te^{F'/T}}{G(F')} dF' = \frac{T}{s} \int_0^x \frac{(f - Tx')}{1 - e^{-x'} + x'^a \Gamma(1-a,x')} dx' \qquad (41)$$

Using this and Eqn. (32), the stationary probability distribution function for the free energy becomes

$$P(F) \sim \frac{x}{1 - e^{-x} + x^a \Gamma(1-a,x)} \exp\left\{\frac{T}{s} \int_0^x \frac{f - Tx'}{1 - e^{-x'} + x'^a \Gamma(1-a,x')} dx'\right\}, x \equiv e^{F/T} \qquad (42)$$

Using this, the average free energy is given by

$$<F> = \int P(F) F dF$$

$$= \frac{\int_0^\infty T dx (\ln x) \frac{1}{1 - e^{-x} + x^a \Gamma(1-a,x)} \exp\left\{\frac{T}{s} \int_0^x dx' \frac{f - Tx'}{1 - e^{-x'} + x'^a \Gamma(1-a,x')}\right\}}{\int_0^\infty dx \frac{1}{1 - e^{-x} + x^a \Gamma(1-a,x)} \exp\left\{\frac{T}{s} \int_0^x dx' \frac{f - Tx'}{1 - e^{-x'} + x'^a \Gamma(1-a,x')}\right\}} \qquad (43)$$

The average number of unzipped base pairs can then be calculated as

$$<X> = \frac{T \int_0^\infty dx (\ln x) \frac{R(x)}{1 - e^{-x} + x^a \Gamma(1-a,x)} \exp\left\{\frac{T}{s} \int_0^x dx' \frac{f - Tx'}{1 - e^{-x'} + x'^a \Gamma(1-a,x')}\right\}}{\int_0^\infty dx \frac{1}{1 - e^{-x} + x^a \Gamma(1-a,x)} \exp\left\{\frac{T}{s} \int_0^x dx' \frac{f - Tx'}{1 - e^{-x'} + x'^a \Gamma(1-a,x')}\right\}}$$

$$- <F> \frac{\int_0^\infty dx \frac{R(x)}{1 - e^{-x} + x^a \Gamma(1-a,x)} \exp\left\{\frac{T}{s} \int_0^x dx' \frac{f - Tx'}{1 - e^{-x'} + x'^a \Gamma(1-a,x')}\right\}}{\int_0^\infty dx \frac{1}{1 - e^{-x} + x^a \Gamma(1-a,x)} \exp\left\{\frac{T}{s} \int_0^x dx' \frac{f - Tx'}{1 - e^{-x'} + x'^a \Gamma(1-a,x')}\right\}} \qquad (44)$$

where

$$R(x) = \frac{T}{s} \int_0^x \frac{dx'}{1 - e^{-x'} + x'^a \Gamma(1-a,x')} \qquad (45)$$

Eqn. (44) can be used to calculate the average number of unzipped base pairs numerically.

In Fig. 1 we show the result of average number of unzipped base pairs $<X>$ as a function of $f$, for different values of the parameter $a < 1$. We can see that for small $f$, $<X>$ increases as $a$ increases, and for all $a < 1$, it is finite. This shows that the DNA double strand is more stable for longer range correlations, in agreement with the finding of Allahverdyan et al [7] and in contrast to the prediction of Lubensky and Nelson [4]. In order to check for any power law dependence of $<X>$ on $f$, we show in Fig. 2 a plot of $-\ln <X>$ versus $\ln f$. Any power law behavior would manifest itself as a straight line. We can see that for small $f$, we have a straight line, but with zero slope.

Our Fokker-Planck equation does not describe the case of completely frozen noise, with $a = 0$. The reason is the following. For the case of completely frozen noise, $h$ is a constant and Eqns. (4) and (5) for the Hamiltonian and free energy can be solved exactly for a particular value of the noise strength

$$H(x) = (x-a)(f+h) \qquad (46)$$

$$Z = \int_a^L dx e^{-b(x-a)(f+h)} = \frac{1}{b(f+h)}[1 - e^{-b(f+h)(L-a)}] \qquad (47)$$

$$-\frac{F}{T} = \ln Z = -\ln[b(f+h)] + \ln\{1 - e^{-b(f+h)(L-a)}\} \qquad (48)$$

For this particular noise strength, the number of unzipped base pairs is given by

$$X = \partial_f F = \frac{T}{f+h} - \frac{L\left(1-\frac{a}{L}\right)e^{-b(f+h)(1-\frac{a}{L})L}}{1 - e^{-b(f+h)L(1-\frac{a}{L})}} \qquad (49)$$

In the limit $L \to \infty$ this becomes

$$X = Lq(-(f+h)) \qquad (50)$$

where $q(x)$ is the Heavyside step function. Averaging this over the noise, we have

$$<X> = L\int_{-\infty}^{\infty} \frac{dh}{\sqrt{2ps}}\exp\left[-\frac{h^2}{2s}\right]q(-(f+h)) = L\int_0^{\infty} \frac{dh}{\sqrt{2ps}}\exp\left[-\frac{h^2}{2s}\right]q(h-f)$$

$$= \frac{L}{2}\left[1 - erf\left(\frac{f}{\sqrt{2s}}\right)\right] \qquad (51)$$

This shows that the average number of unzipped base-pairs $<X>$ depends explicitly on L in the whole range of physical parameters. A similar result was obtained in ref. [7] by a more elaborate calculation. This means that a Fokker-Planck approach cannot be used to study the case of frozen noise because in the Fokker-Planck equation the thermodynamic limit $L \to \infty$ has already been taken. There is therefore no dependence on L in the stationary free energy distribution $P(F)$ and consequently also no dependence on L in the quantity $<X>$ obtained using this distribution and Eqn. (33) and (34). That the thermodynamic limit does not exist for $<X>$ in the case of completely frozen noise is probably related to the unphysical nature of the completely frozen noise itself. We are assuming that for long range noise with $1 > a > 0$, but not completely frozen, the thermodynamic limit for $<X>$ exists and is independent of L. Otherwise a Fokker-Planck approach cannot be applied at all.

## VII. Conclusion

In this paper we have presented a detailed study of a model of double-stranded DNA being unzipped by an external force. The model was originally proposed by Allahverdyan et al [7] and is a simplified version of that proposed by Lubensky and Nelson [4], in that it neglects the restoring force due to the single strands after the DNA is partially unzipped. The neglect of configurational entropy contribution of the DNA and also the bubble contribution in the model make it impossible to derive the qualitative correct features of the phase diagram in the temperature-force plane [1-3]. Even then, the model reproduces the behavior of the average number of unzipped base pairs $<X> \sim 1/f^2$ of the Lubensky-Nelson model in the white noise limit. The simplification in the model facilitates its study and hopefully will produce an answer on the effect of long range correlation on unzipping. Still there has been very little work done on this latter problem, besides the heuristic argument of Lubensky and Nelson [4] and the numerical simulation of Allahverdyan et al. [7]. Our Fokker-Planck equation approach may provide a new perspective.

We derive and solve a Fokker-Planck equation for the stationary distribution of the free energy for a model in which the autocorrelation function of the random DNA sequence can be a general form, including long range correlations. The only approximation we have made in the derivation is that shown in Eqn. (27). In the case of Orstein-Uhlenbeck noise, characterized by a finite correlation length, our result reduces to the exact result of Allahverdyan et al, with the average number of unzipped base pairs going as $<X> \sim 1/f^2$ in the white noise limit, where $f$ is the deviation from the critical force. In the case of long rang correlated noise, where the integrated autocorrelation is divergent, we find that $<X>$ is finite at $f = 0$, with its value decreasing as the correlations become longer range. This shows that long range correlations actually stabilize the DNA sequence against unzipping. Our result is also in agreement with the findings of Allahverdyan et al [7], obtained using numerical generation of the correlated noise, but contradicts the prediction of Lubensky and Nelson [4] $<X> \sim 1/f^{2/a}, a < 1$. Our result for the long range correlated noise is based on the assumption that for long range noise with $1 > a > 0$, but not completely frozen, the thermodynamic limit for the average unzipped base-pair $<X>$ exists and is independent of L. Otherwise a Fokker-Planck approach cannot be applied at all. Such is the case for the completely frozen noise.

**Acknowledgement.** Research supported by NSF Epscor LASIGMA and the Louisiana Board of Regents Support Fund Contract Number LEQSF(2007-10)-RD-A-29. PML would like to thank the hospitality of the Institute of Theoretical Physics, Chinese Academy of Sciences, where part of this work was carried out.

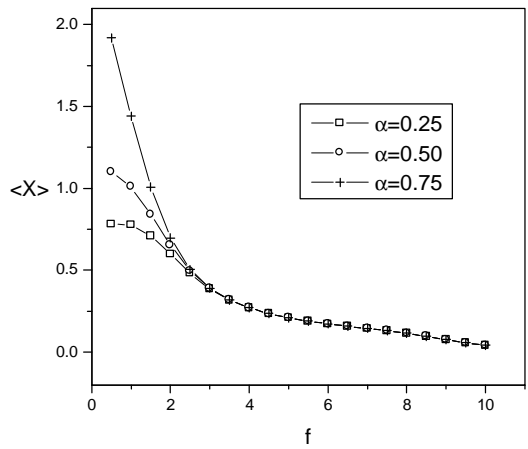

Fig. 1: Average number of unzipped base pairs as function of $f$, for different values of $a$.

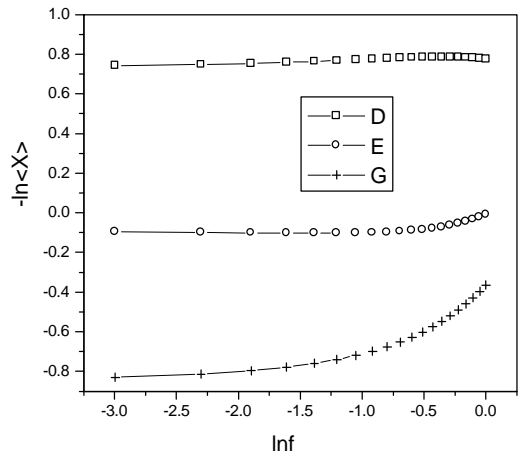

Fig. 2: $-\ln <X>$ versus $\ln f$, for different values of $a$.